\documentclass[journal]{IEEEtran}
\usepackage{stackrel}
\usepackage{epsf}
\usepackage{amsmath,amssymb}
\usepackage{graphicx}
\usepackage{color}
\usepackage{cite}
\usepackage{ifthen}
\usepackage{float}
\usepackage{subfig}
\usepackage{times}
\usepackage{graphicx}
\usepackage{color}
\usepackage{cite}
\usepackage{bm}
\usepackage{stackrel}
\usepackage[linesnumbered,ruled]{algorithm2e}

\newtheorem{lemma}{{\bf Lemma}}

\newtheorem{result}{{\bf Result}}

\newenvironment{proof}[1][Proof]{\begin{trivlist}}

\newcommand{\beq}{\begin{equation}}
\newcommand{\eeq}{\end{equation}}
\newcommand{\barr}{\begin{array}}
\newcommand{\earr}{\end{array}}

\newcommand{\benum}{\begin{enumerate}}
\newcommand{\eenum}{\end{enumerate}}

\newcommand{\bit}{\begin{itemize}}
\newcommand{\eit}{\end{itemize}}

\newcommand{\bc}{\begin{center}}
\newcommand{\ec}{\end{center}}

\newcommand{\bdes}{\begin{description}}
\newcommand{\edes}{\end{description}}

\newcommand{\bfig}{\begin{figure}}
\newcommand{\efig}{\end{figure}}

\newcommand{\bemq}{\begin{quote} \begin{em}}
\newcommand{\eemq}{\end{em} \end{quote}}

\newcommand{\bmp}{\begin{minipage}}
\newcommand{\emp}{\end{minipage}}

\newcommand{\brac}[1]{\left({#1}\right)}
\newcommand{\sbrac}[1]{\left[{#1}\right]}
\newcommand{\cbrac}[1]{\left\{{#1}\right\}}

\newcommand{\supth}{^{{\mathrm{th}}}}

\newcommand{\expect}[1]{{\mathbb E}\left[{#1}\right]}

\IEEEoverridecommandlockouts
\begin{document}

\title{Power Allocation for Massive MIMO-based, Fronthaul-constrained Cloud RAN Systems}

\author{Jobin Francis and Gerhard Fettweis, {\it Fellow, IEEE}
\thanks{The authors are with the Vodafone Chair Mobile Communication Systems at Technische Universit{\"a}t Dresden (TUD), Germany.}
\thanks{Emails: jobin.francis@tu-dresden.de and gerhard.fettweis@tu-dresden.de.}
}

\IEEEaftertitletext{\vspace{-0.5\baselineskip}}
\maketitle

\begin{abstract}
Cloud radio access network (C-RAN) and massive multiple-input-multiple-output (MIMO) are two key enabling technologies to meet the diverse and stringent requirements of the 5G use cases. In a C-RAN system with massive MIMO, fronthaul is often the bottleneck due to its finite capacity and transmit precoding is moved to the remote radio head to reduce the capacity requirements on fronthaul. For such a system, we optimize the power allocated to the users to maximize first the weighted sum rate and then the energy efficiency (EE) while explicitly incorporating the capacity constraints on fronthaul. We consider two different fronthaul constraints, which model capacity constraints on different parts of the fronthaul network. We develop successive convex approximation algorithms that achieve a stationary point of these non-convex problems. To this end, we first present novel, locally tight bounds for the user rate expression. They are used to obtain convex approximations of the original non-convex problems, which are then solved by solving their dual problems. In EE maximization, we also employ the Dinkelbach algorithm to handle the fractional form of the objective function. Numerical results show that the proposed algorithms significantly improve the network performance compared to a case with no power control and achieves a better performance than an existing algorithm.
\end{abstract}


\section{Introduction}
\label{sec:Intro}

The fifth generation (5G) cellular standard, in addition to enhancing the mobile broadband experience, is envisioned to enable new use cases such as autonomous vehicle control, factory automation, virtual and augmented reality, remote medicine, and internet-of-things. These use cases place diverse and stringent requirements on the cellular network such as higher data rates, energy efficiency (EE), reliability and availability, and lower latency~\cite{tr:itu5G_reqs}. Two promising technologies to meet these requirements while reducing the capital and operational expenditures are cloud radio access network (C-RAN) and massive multiple-input-multiple-output (MIMO). 

C-RAN is a novel network architecture in which the radio frequency processing is done at remote radio heads (RRHs) while the baseband processing for all the cells is done centrally at the baseband unit (BBU) pool. The transport network connecting the BBU pool to the RRHs is referred to as \emph{fronthaul}. Advantages of C-RAN include higher spectral efficiency (SE) and EE, better cell-edge performance and mobility management. These gains are possible through centralized/cooperative baseband processing. C-RAN also facilitates antenna site simplification and enables pooling gains by sharing the co-located computing resources of multiple cells.

In massive MIMO, the base stations are equipped with a very large number of antennas. Compared to regular MIMO systems, it significantly improves the SE, EE, and coverage through higher spatial multiplexing and array gains. Further, it is quite robust to hardware impairments enabling the use of low cost, low precision components. To avail these benefits in a C-RAN system, massive MIMO-based RRHs are used.

Despite the many advantages of C-RAN, realizing them in practice is challenging due to non-ideal fronthaul. Often, fronthaul is capacity or latency constrained. This can limit the potential gains of C-RAN given the stringent requirements it places on fronthaul. For example, the fronthaul capacity required to transport a $20$~MHz long term evolution (LTE) signal for $2 \times 2$ MIMO, which gives a peak downlink data rate of $150$~Mbps, is $2.46$~Gbps~\cite{Wubben_SPM_2014}. Further, it increases linearly as the number of antennas increases. Thus, multiple optical fiber links might be needed for a massive MIMO-based RRH, which is not scalable and economically viable. 



Functional splitting has been proposed to reduce the fronthaul requirements~\cite{Wubben_SPM_2014, Peng_Survey_2016}. In it, some baseband functionalities are offloaded to the RRH -- referred to now as remote radio unit (RRU). As more functionalities are moved to the RRU, the capacity and latency requirements on fronthaul decreases. However, this is at the cost of reduced centralization gains. For a C-RAN system with massive MIMO-based RRUs, a suggested functional split involves moving transmit precoding to the RRUs~\cite{Jens_Eurasip_2017}. Thereby, the traffic in fronthaul is dependent only on users' data rates and not on the number of antennas. The centralization gains can still be obtained by joint power control and coordinated scheduling between multiple cells.

In this paper, we optimize the allocation of transmit powers to the users to maximize weighted sum rate~(WSR) and EE, which are two popularly used network performance measures. Since fronthaul is often the bottleneck in C-RAN, the optimization problems explicitly model the constraint on fronthaul capacity. This makes the problem challenging and different from most of the existing techniques for C-RAN design.  The power control can be done at a slower time-scale leveraging the fact that the user rates can be accurately approximated in terms of their large-scale fading coefficients owing to the channel hardening effect in massive MIMO~\cite{Bjornson_TWC_2016}. Further, coordinated scheduling is implicitly done via power control as a user is not scheduled if the power allocated to it is zero. 

\subsection{Related Literature}
\label{sec:RelLit}

We first discuss prior works on sum rate maximization that explicitly incorporates fronthaul capacity constraints and then on EE maximization. The transmit precoding and compression covariance matrices are jointly optimized in~\cite{Shamai_TWL_2014} to maximize sum rate by using a successive convex approximation (SCA) algorithm. Stochastic optimization problems to maximize the throughput of a C-RAN system with dynamic traffic and time-varying channels are studied in~\cite{Peng_Access_2017} while ensuring network stability. In~\cite{Niu_TWC_2016}, channel allocation and user association are optimized to maximize WSR of a sliced network while ensuring isolation between slices via interference constraints. Transmit beamformers and powers are optimized to maximize WSR in~\cite{Quek_ICC_2016} while guaranteeing minimum signal-to-interference-plus-noise ratio (SINR) for the users. However, the models and optimization problems considered in~\cite{Shamai_TWL_2014, Peng_Access_2017, Niu_TWC_2016, Quek_ICC_2016} differ from ours. Specifically, fronthaul compression instead of functional splitting is considered~\cite{Shamai_TWL_2014} to lower the fronthaul data rate, no power optimization is considered in~\cite{Niu_TWC_2016}, dynamic traffic is considered in~\cite{Peng_Access_2017}, and per-user fronthaul constraints, which reduce to simpler SINR constraints, are considered in~\cite{Quek_ICC_2016}.



In~\cite{Luong_TSP_2017}, transmit beamforming, RRH selection, and user association are jointly optimized to maximize the weighted difference between sum rate and power expended. However, this optimization objective is different from ours and the reformulations of the optimization problem used in it do not extend to WSR maximization. In~\cite{Juntti_ICASSP_2017}, sum rate is maximized by optimizing transmit precoding and user association subject to SINR constraints. Again, the reformulations used in it do not extend to our setting, as has also been noted in~\cite{WeiYu_ICC_2015}. Transmit precoding is optimized in~\cite{WeiYu_Access_2014} to maximize WSR by using the weighted minimum mean square error (WMMSE) algorithm~\cite{Shi_TSP_2011}. The fronthaul constraint is handled by reformulating it as a weighted power constraint. However, this is ad hoc. Hence, no guarantees on the convergence is provided in~\cite{WeiYu_Access_2014} or the point to which it converges, if at all.

EE is maximized in~\cite{Juntti_ICC_2017} by optimizing transmit precoding, compression covariance matrices, and user association. In~\cite{Peng_TMT_2016}, EE is maximized in a dynamic traffic scenario while ensuring the stability of the queues. However, the models and the optimization problems considered in these works are quite different from ours as they focus on systems with fronthaul compression and dynamic traffic, respectively. Transmit beamforming, RRH selection, and user association are jointly optimized in~\cite{Luong_ICC_2017} to maximize EE. An SCA algorithm is developed, where a sequence of computationally intensive mixed integer second order cone programs are solved. In~\cite{Schober_TWC_2012}, transmit powers and subcarrier allocation in an orthogonal frequency division multiple access system is optimized to maximize EE. It uses the Dinkelbach's (DB) algorithm and duality to solve the problem. However, the the duality gap is not zero. The algorithm in it also lacks convergence guarantees as global solutions are not ensured in each DB iteration. 

We also note that none of the above works consider massive MIMO, which has a different functional form for SINR due to the interference resulting from pilot contamination. We now discuss prior works on resource allocation in massive MIMO systems. Power allocation to maximize sum rate and EE is considered in~\cite{Zhang_TWC_2015, Nguyen_Access_2015} and~\cite{Alessio_EE_Book}, respectively. In~\cite{Bjornson_TWC_2015}, transmit powers, number of users and number of antennas are optimized to maximize EE. Logarithmic utility is maximized in~\cite{QYe_TCom_2016} by optimizing user association and resource blanking in a heterogeneous network (HetNet) with massive MIMO-based macro base stations and small cells. The trade-off between SE and EE is studied in~\cite{Hao_ComLet_2016} by optimizing user association and transmit powers. However, the works in~\cite{Zhang_TWC_2015, Nguyen_Access_2015, Alessio_EE_Book, Bjornson_TWC_2015, QYe_TCom_2016, Hao_ComLet_2016} do not consider fronthaul capacity constraints. A HetNet similar to that in~\cite{QYe_TCom_2016} is considered in~\cite{Bhargava_TWC_2016, Liu_WComLet_2016} with wireless backhaul to the macro for the the small cells. For this system, user association and backhaul bandwidth allocation are optimized to maximize sum rate in~\cite{Liu_WComLet_2016} and logarithmic utility in~\cite{Bhargava_TWC_2016}. However, this system model is very different from ours. The WSR of a C-RAN system with massive MIMO-based RRUs is maximized in~\cite{Rajesh_TVT_2017} by optimizing user association, fronthaul link selection, power allocation, and the number of antennas serving a user. However, it considers a simpler fronthaul constraint that the
number of users served over a fronthaul link is below a predefined
value instead of a capacity constraint.



\subsection{Contributions}
\label{sec:Contr}

We present an SCA-based optimization framework to coordinate the transmit powers of users in a C-RAN system, where RRUs are equipped with massive MIMO and fronthaul is capacity limited.  Our specific contributions are as follows.

\emph{WSR Maximization:} We first focus on maximizing the WSR of the network. In order to apply the SCA algorithm, we develop a novel upper bound and a lower bound for the non-convex user rate function. Both these bounds are needed as user rates appear in the objective function and in the fronthaul capacity constraint. These bounds are used to obtain a convex approximation to the non-convex WSR maximzation problem, which is then solved in each SCA iteration via dual decomposition. The SCA algorithm  is guaranteed to monotonically converge to a point satisfying the Karush-Kuhn-Tucker (KKT) conditions of the original problem.

We consider two different fronthaul capacity constraints. We first consider the case where the individual fronthaul links to each RRU is capacity-constrained. Then, the fronthaul link that carries the data for all the RRUs is considered to be constrained. The proposed optimization framework can handle both these constraints.

\emph{EE Maximization:} We then focus on maximizing the EE of the network. We first propose a novel lower bound on EE in order to apply the SCA algorithm. The convex approximation to the non-convex EE maximization problem is solved in each SCA iteration using the DB algorithm and dual decomposition. As before, the proposed SCA algorithm is guaranteed to converge to a KKT point. We also consider two different fronthaul capacity constraints and develop algorithms for them.

\emph{Performance Benchmarking:} We then evaluate the performance of the proposed algorithms for maximal ratio transmission (MRT) and zero-forcing (ZF) precoding schemes and for two different user association rules, namely, distance-based and signal power-based association. For WSR maximization, we benchmark the proposed algorithms against the WMMSE algorithm in~\cite{WeiYu_Access_2014} and a baseline scheme with no power control. For EE maximization, we benchmark against an extension of the WMMSE algorithm in~\cite{WeiYu_Access_2014} and the baseline scheme. The proposed algorithms significantly improve WSR and EE compared to the baseline scheme. They have a similar or better performance the WMMSE algorithm and its extension. 
 
\subsection{Organization and Notations}

The paper is organized as follows. Section~\ref{sec:SysMod} describes the system model. Section~\ref{sec:WSRMax} and \ref{sec:WEEMax} develop algorithms for WSR and EE maximization, respectively. Numerical results are presented in Section~\ref{sec:NumRes}. Our conclusions follow in Section~\ref{sec:Concls}.

We use boldface lowercase and uppercase letters to denote vectors and matrices, respectively. We denote a vector of zeros and ones by $\mathbf{0}$ and $\mathbf{1}$, respectively. Further, $\mathbf{x}^H$, $||\mathbf{x}||$ and $\mathbf{x} \geq \mathbf{0}$ respectively denote the hermitian, the Euclidean norm, and that each element of $\mathbf{x}$ is non-negative. An $n\times n$ identity matrix is denoted by $\mathbf{I}_n$. Define $[a]^+=\max\cbrac{0,a}$. The expectation operator is denoted by $\expect{\cdot}$, gradient of a function $f$ by $\nabla f$, and indicator of an event $\mathcal{A}$ by $I_{\mathcal{A}}$.

\section{System Model}
\label{sec:SysMod}

\begin{figure}
\centering
\includegraphics[width=\linewidth]{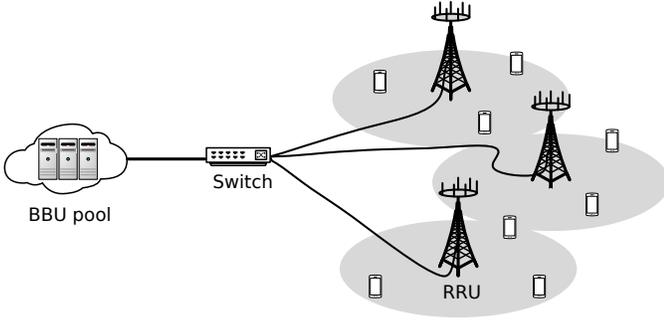}
\caption{A C-RAN system with fronthaul connecting BBU pool to massive MIMO-based RRUs.}
\label{fig:SysMod}
\end{figure}

The C-RAN system consists of a BBU pool, $L$ RRUs and $K$ users as shown in Figure~\ref{fig:SysMod}. We consider a fronthaul network with a switch that demultiplexes the data from BBU to the RRUs. Since the  fronthaul link between the BBU and switch carries the data for all the RRUs, it is designed to have a higher capacity than the links between the switch and RRUs. The users are equipped with a single antenna while RRUs are equipped with $N \gg 1$ antennas. The users associate statically and uniquely to the RRUs. Let $\mathcal{K}_l$ and $j_k$ respectively denote the set of users associated to RRU $l$, for $l=1,\ldots,L$, and the RRU to which user $k$ is associated to, i.e., $j_k=l$, for $k\in \mathcal{K}_l$. We shall consider different association rules in Section~\ref{sec:NumRes}.

The system operates in time division duplex mode and channel reciprocity is assumed. We consider a coherence interval of $T_c$ symbols and $T_p < T_c$ symbols are used for uplink training. Then, $\tau= \kappa(1-T_p/T_c)$ is the fraction of symbols available for downlink data transmission, where $0\leq \kappa \leq 1$ determines the partitioning of data symbols between uplink and downlink. Let $\mathbf{h}_{jk} \in \mathcal{C}^N$ denote the complex baseband channel gain vector between RRU $j$ and user $k$. It is modeled as a zero mean complex Gaussian random vector with covariance matrix $\beta_{jk}\mathbf{I}_N$, where $\beta_{jk}$ is the large-scale fading coefficient that models pathloss and shadowing\footnote{Although we have ignored spatial correlation, extension to the case with spatial correlation is possible through deterministic equivalents~\cite{Hoydis_JSAC_2013}.}. This assumption allows for tractable, closed-form expressions for SINR and has also been assumed in~\cite{Bjornson_TWC_2016}.

 
\subsection{Uplink Channel Estimation}

In order to estimate the uplink channel, users transmit orthogonal pilot sequences of length $T_p$ to the RRUs. The pilot sequences are shared by the users since the number of pilot sequences is typically less than the number of users. Let $b_{k}$ denote the pilot sequence used by user $k$ and $\mathcal{B}_{k}$ denote the set of users that use the pilot sequence $b_k$. As shown in~\cite{Bjornson_TWC_2016}, the minimum mean square estimate $\hat{\mathbf{h}}_{jk}$ of $\mathbf{h}_{jk}$ is given by $\hat{\mathbf{h}}_{jk}=\beta_{jk}\tilde{\mathbf{h}}_{jb_{k}}$, where $\tilde{\mathbf{h}}_{jb_{k}}$ is as follows:
\begin{equation}
\label{Eq:ChanEst}
\tilde{\mathbf{h}}_{jb_{k}}=\frac{\sum_{i\in \mathcal{B}_{k}} \mathbf{h}_{ji}+ \sqrt{\frac{\sigma^2}{T_pP_{\text{tr}}}} \mathbf{n}_{jb_{k}}}{\sum_{i\in \mathcal{B}_{k}} \beta_{ji}+{\sigma^2}/\brac{T_pP_{\text{tr}}}},
\end{equation}  
where $\mathbf{n}_{jb_{k}}$ is an independent and identical complex Gaussian random vector of unit variance, $\sigma^2$ is the thermal noise power, and $P_{\text{tr}}$ is the uplink transmit power. Further, $\hat{\mathbf{h}}_{jk}$ is a complex Gaussian random vector with covariance matrix $\theta_{jk}\mathbf{I}_N$, where $\theta_{jk}={\beta^2_{jk}}/\brac{\sum_{i\in \mathcal{B}_{k}} \beta_{ji}+{\sigma^2}/\brac{T_pP_{\text{tr}}}}$.

Notice that the channel estimates for users that use the same pilot sequence differ only by a scaling factor. This is referred to as \emph{pilot contamination}. It places a fundamental limit on the channel estimation accuracy and, in turn, the performance of massive MIMO.

\subsection{Downlink Data Transmission}

We consider linear transmit strategies in order to keep the encoding process simple. The estimated uplink channel gains are used to generate the downlink precoding vectors. Let $\mathbf{f}_{k} \in \mathbb{C}^N$ and $p_{k}$ respectively denote the precoder and power applied by RRU $j$ to the data $s_{k}$ intended for user $k$. Then, the transmitted signal $\mathbf{z}_j$ of RRU $j$ is $
\mathbf{z}_j=\sum_{k \in \mathcal{K}_j} s_{k} p_{k} \mathbf{f}_{k}$.

The power of the transmit signal $\mathbf{z}_j$ should be below the power budget $P_t$ of RRU $j$. Without loss of generality, we assume the power budget to be the same for all the RRUs. The users' data are uncorrelated and are normalized such that $\expect{|s_{k}|^2}=1$, for all $k \in \mathcal{K}_j$. Then, the power constraint for RRU $j$ can be expressed as
\begin{equation}
\label{Eq:PowConstr1}
\expect{\mathbf{z}_j^H \mathbf{z}_j} = \sum_{k \in \mathcal{K}_j} p_{k} \expect{\mathbf{f}_{k}^H \mathbf{f}_{k}} \leq P_t,\; \text{for}\; j=1,\ldots,L.
\end{equation}

The received signal $y_{k}$ at user ${k}$ is $ y_{k}=\sum_{j=1}^L \mathbf{h}_{jk}^H \mathbf{z}_j + n_{k}$, where $n_{k}$ is circularly symmetric complex Gaussian noise with variance $\sigma^2$. An achievable rate $R_{k}$ when user $k$ treats interference as noise is $R_{k}= \tau \log_2\brac{1+\gamma_{k}}$, where $\gamma_{k}$ denotes the SINR of user $k$. It is given by~\cite{Bjornson_TWC_2016}
\begin{equation}
\label{Eq:SINRGen}
\gamma_{k}=\frac{p_{k} \left\lvert \expect{\mathbf{h}_{j_k k}^H\mathbf{f}_{k}} \right\rvert^2}{\sigma^2+\sum_{i=1}^K p_{i} \expect{\left\lvert \mathbf{h}_{j_ik}^H\mathbf{f}_{i}\right\rvert^2} - p_{k} \left\lvert \expect{\mathbf{h}_{j_k k}^H\mathbf{f}_{k}} \right\rvert^2}.
\end{equation}

We consider MRT and ZF precoding schemes. Their precoding vectors for user $k$ are as follows:
\begin{equation}
\label{Eq:PrecodVec}
\mathbf{f}_{k} = \left\{\begin{array}{lr}
        \frac{\hat{\mathbf{h}}_{k}}{\sqrt{\expect{||\hat{\mathbf{h}}_{k}||^2}}}, & \text{for MRT}, \\
        \frac{\mathbf{G}_{l}\mathbf{g}_{b_{k}}}{\sqrt{\expect{||\mathbf{G}_{l}\mathbf{g}_{b_{k}}||^2}}}, & \text{for ZF},
\end{array}\right.
\end{equation}
where $\mathbf{g}_{b_{k}}$ is the $b_{k}\supth$ column of the matrix $\brac{\mathbf{G}_{l}^H\mathbf{G}_{l}}^{-1}$ with $\mathbf{G}_{l}=[\tilde{\mathbf{h}}_{l1},\ldots,\tilde{\mathbf{h}}_{lT_p}]$. Under perfect channel estimation, this ZF precoding scheme cancels out interference even to users not associated with RRU $l$.

The SINR expression in~\eqref{Eq:SINRGen} can be evaluated in closed-form for MRT and ZF precoders. They have a similar form and is given by~\cite{Bjornson_TWC_2016}:
\begin{equation}
\label{Eq:SinrExpnCompct}
\gamma_{k} = \frac{v p_{k}\theta_{j_kk}}{\sum_{i=1}^K  p_{i} \brac{w_{j_ik}+v\theta_{j_ik} s_{ik}} - v p_{k}\theta_{j_kk} + \sigma^2},
\end{equation}
where $s_{ik}=1$ if $i \in \mathcal{B}_{k}$ and $0$ otherwise. Here, $v$ and $w_{j_ik}$ depend on the precoding scheme. For MRT, $v=N$ and $w_{j_ik}=\beta_{j_ik}$, and for ZF, $v=N-T_p$ and $w_{j_ik}=\beta_{j_ik}-\theta_{j_ik}$. 

The SINR expression in \eqref{Eq:SinrExpnCompct} is quite intuitive. The numerator term represents the received signal power while $\sum_{i=1}^K  v\theta_{j_ik} s_{ik}p_{i}$ and $\sum_{i=1}^K  w_{j_ik}p_{i}$ in the denominator represent the interference due to pilot contamination and the multi-user interference, respectively. Note that the interference due to pilot contamination  increases with $N$ similar to the numerator. Thus, pilot contamination becomes the limiting factor asymptotically in $N$. Comparing the SINR expressions of MRT and ZF, we see that the array gain is higher for MRT while the multi-user interference is lower for ZF. This is because ZF attempts to nullify the multi-user interference, while MRT tries to maximize the signal-to-noise-ratio. Further, since the SINR expression in \eqref{Eq:SinrExpnCompct} depends only on the large-scale fading coefficients $\beta_{jk}$ and not on the instantaneous channel realizations $\mathbf{h}_{jk}$, power control needs to be carried out only at a slower time scale. This reduces the system complexity and the computational requirements at the BBU. 

\subsection{Fronthaul Constraints} 

We now model the capacity constraints on fronthaul. The constraint differs depending on which part of fronthaul is the bottleneck. If the links between RRUs and switch are the constraints, then the data rate served by each RRU should be less than the link capacity $C_{\text{plc}}$\footnote{Without loss of generality, we set $C_{\text{plc}}$ to be the same for all the fronthaul links between switch and RRUs.}. Thus, we have
\begin{equation}
\label{Eq:PLC1}
\tau\sum_{k \in \mathcal{K}_l} \log_2\brac{1+\gamma_{k}} \leq \eta C_{\text{plc}},\; \text{for}\; l=1,\ldots,L,
\end{equation}
where $\eta$ is the ratio of fronthaul bandwidth to the downlink bandwidth. We shall refer to it as \emph{per-link capacity constraints}. 

On the other hand, if the link between BBU and switch is the constraint, then the aggregate data rate served by all the RRUs should be less than the link capacity $C_{\text{sc}}$. Thus, we have
\begin{equation}
\label{Eq:SC}
\tau\sum_{l=1}^L \sum_{k \in \mathcal{K}_l} \log_2\brac{1+\gamma_{k}} \leq \eta C_{\text{sc}}.
\end{equation}
We shall refer to it as \emph{sum-capacity constraint}.

\subsection{Power Consumption Model}
\label{sec:PowCons}

The power consumed in a C-RAN system has static and dynamic power components. The static component $P_S$ involves the power consumed by users during uplink training and data transmission, the circuit power in RRUs, and the power expended in fronthaul. The power consumed by a user during uplink training and data transmission is $P_{\text{UE}}=(1-\tau)P_{\text{tr}}/\omega_\text{UE}$, where $\omega_\text{UE}$ is the power amplifier efficiency of the user equipment. As per~\cite{Bjornson_TWC_2015}, the circuit power in massive MIMO is given by $\varrho+N\varsigma$. Here, $\varsigma$ models the power components that scales with the number of antennas, which involves the power consumed by converters, mixers, and filters in each antenna branch and $\varrho$ models the power components that do not scale with the number of antennas. The power expended in fronthaul $P_\text{FH}$ is assumed to be fixed as has been assumed in~\cite{Schober_TWC_2012, Peng_TMT_2016}. This is justified as fronthaul is never idle and $P_\text{FH}$ does not depend on the precoding vectors as has been assumed in~\cite{He_TCOM_2013, Juntti_ICC_2017} since precoding is done at RRU. Thus, the static component is $P_S= KP_{\text{UE}} + L \brac{\varrho+N\varsigma}+P_\text{FH}$.

The dynamic component depends on the signal transmitted in the downlink and is given by $\brac{\tau/\omega_\text{RRU}} \sum_{i=1}^K P_i$, where $\omega_\text{RRU}$ is the power amplifier efficiency of RRUs. Therefore, the total power consumed is $P_{\text{C-RAN}}=P_S+\brac{\tau/\omega_\text{RRU}} \sum_{i=1}^K P_i$.

\section{WSR Maximization}
\label{sec:WSRMax}

In this section, we optimize the transmit powers to maximize the WSR. The weights can be assigned to prioritize users. We first formulate the optimization problem and develop an SCA algorithm for per-link capacity constraint and then for sum-capacity constraint.

\subsection{Problem Formulation}
Let $\bm{p}=[p_{1},\ldots,p_K]$. The WSR maximization subject to power and per-link capacity constraints is as follows:
\begin{align}
\label{Eq:ObjFnWsr}
\mathbf{P}1:\; \max_{\bm{p}\geq \mathbf{0}} \;& \sum_{l=1}^L \sum_{k \in \mathcal{K}_l} \alpha_{k} \tau \log_2\brac{1+\gamma_{k}}, \\
\label{Eq:PLC}
\text{s.t.} \; & \sum_{k \in \mathcal{K}_l} \log\brac{1+\gamma_{k}} \leq \tilde{\eta} C_{\text{plc}}, \\
\label{Eq:PowConstr}
 & \sum_{k \in \mathcal{K}_l} p_{k} \leq P_t, \; \text{for}\; l=1,\ldots,L,
\end{align}
where $\tilde{\eta}\!=\!\eta\log(2)/\tau$ and $\alpha_{k}$ is the weight assigned to user $k$.

The WSR maximization problem is a non-convex, strongly NP-hard problem even without the non-convex fronthaul  constraints~\cite{Luo_JSTSP_2008}. Therefore, finding globally optimal solutions is quite challenging. We develop a lower computational complexity, iterative algorithm to find a locally optimum solution via SCA. A brief introduction to SCA is given in Appendix~\ref{app:SCA}.

\subsection{Proposed SCA Algorithm}

In order to apply the SCA algorithm, we need a locally tight lower bound and an upper bound, which are convex functions of $\bm{p}$, for the non-convex function $\log\brac{1+\gamma_{k}}$ as it appears in the objective function in \eqref{Eq:ObjFnWsr} and constraints in $\eqref{Eq:PLC}$. For this, we have the following key result.

\begin{result}
\label{res:CvxApx}
For any two power vectors $\bm{p}$ and $\bm{p}^{(r)}$, we have $\log\brac{1+\gamma_{k}} \geq G_{k}\brac{\bm{p};\bm{p}^{(r)}}$ and $\log\brac{1+\gamma_{k}} \leq H_{k}\brac{\bm{p};\bm{p}^{(r)}}$, where $G_{k}\brac{\bm{p};\bm{p}^{(r)}}$ and $H_{k}\brac{\bm{p};\bm{p}^{(r)}}$ are given by
\begin{align}
\nonumber
&G_{k}\brac{\bm{p};\bm{p}^{(r)}}  =  \log\brac{1+\gamma_{k}^{(r)}} \\ \nonumber
&\hspace{0.4cm}- \sum_{i=1}^K \left[  \frac{\brac{p_i-p_i^{(r)}}\brac{w_{j_ik}+v\theta_{j_ik} s_{ik} I_{\{i\neq k\}}}}{\sum_{n=1}^K  p_{n}^{(r)} \brac{w_{j_nk}+v\theta_{j_nk} s_{nk}} - v p_{k}^{(r)}\theta_{lk} + \sigma^2} \right. \\ \label{Eq:LB}
&\hspace{0.4cm} \left. - \frac{\brac{\log(p_i)-\log(p_i^{(r)})}\brac{w_{j_ik}+v\theta_{j_ik} s_{ik}}}{\sum_{n=1}^K  p_{n}^{(r)} \brac{w_{j_nk}+v\theta_{j_nk} s_{nk}} + \sigma^2} \right],\\ \nonumber
& H_{k}\brac{\bm{p};\bm{p}^{(r)}} = \log\brac{1+\gamma_{k}^{(r)}} \\ \nonumber
& \hspace{0.4cm} + \sum_{i=1}^K \left[ \frac{\brac{p_i-p_i^{(r)}}\brac{w_{j_ik}+v\theta_{j_ik} s_{ik}}}{\sum_{n=1}^K  p_{n}^{(r)} \brac{w_{j_nk}+v\theta_{j_nk} s_{nk}} + \sigma^2} \right. \\ \label{Eq:UB}
& \hspace{0.4cm} \left. -  \frac{\brac{\log(p_i)-\log(p_i^{(r)})}\brac{w_{j_ik}+v\theta_{j_ik} s_{ik} I_{\{i\neq k\}}}}{\sum_{n=1}^K  p_{n}^{(r)} \brac{w_{j_nk}+v\theta_{j_nk} s_{nk}} - v p_{k}^{(r)}\theta_{lk} + \sigma^2} \right], \!\!
\end{align}
where $\gamma_{k}^{(r)}$ is SINR of user $k$ at $\bm{p}^{(r)}$. Further, $G_{k}\brac{\bm{p};\bm{p}^{(r)}}$ and $H_{k}\brac{\bm{p};\bm{p}^{(r)}}$ satisfy the properties in Lemma~\ref{Lem:SCA}.
\end{result}
\begin{proof}
The proof is relegated to Appendix~\ref{app:Bounds}.
\end{proof}

We note that these bounds differ from those used in~\cite{Wang_ICC_2012, Shamai_TWL_2014, Luong_TSP_2017} and yield simpler convex problems as we show below. Using the above bounds, an equivalent convex approximation $\mathbf{P}2$ to $\mathbf{P}1$ at the point $\bm{p}^{(r)}$ is obtained in the $r\supth$ iteration of the SCA algorithm. It is given by
\begin{align}
\label{Eq:ObjFnCvx}
\mathbf{P}2:\; \min_{\bm{p}\geq \mathbf{0}} & -\sum_{l=1}^L \sum_{k \in \mathcal{K}_l} \alpha_{k} G_{k}\brac{\bm{p};\bm{p}^{(r)}} \; \text{s.t.} \; \eqref{Eq:PowConstr}, \\ \label{Eq:PLCCvx}
 & \sum_{k \in \mathcal{K}_l} H_{k}\brac{\bm{p};\bm{p}^{(r)}} \leq \tilde{\eta} C_{\text{plc}}, \; l=1,\ldots,L. 
\end{align}
It can be solved by the Lagrange multipliers method. Let $\bm{\lambda}=[\lambda_1,\ldots,\lambda_L] \geq \mathbf{0}$ denote the Lagrange multipliers associated with the fronthaul constraints in \eqref{Eq:PLCCvx}. Then, the Lagrange function $\mathcal{L}\brac{\bm{p},\bm{\lambda};\bm{p}^{(r)}}$ is given by
\begin{multline}
\label{Eq:Lang1}
\mathcal{L}\brac{\bm{p},\bm{\lambda};\bm{p}^{(r)}} = -\sum_{l=1}^L \sum_{k \in \mathcal{K}_l} \alpha_{k} G_{k}\brac{\bm{p};\bm{p}^{(r)}} \\
 + \sum_{l=1}^L \lambda_l \brac{\sum_{k \in \mathcal{K}_l} H_{k}\brac{\bm{p};\bm{p}^{(r)}} - \tilde{\eta} C_{\text{plc}}}.
\end{multline}
Then, the dual function is $\mathcal{G}\brac{\bm{\lambda};\bm{p}^{(r)}}=\min_{\mathcal{P}} \mathcal{L}\brac{\bm{p},\bm{\lambda};\bm{p}^{(r)}}$, where $\mathcal{P}=\cbrac{\bm{p}\geq \mathbf{0}, \sum_{k \in \mathcal{K}_l} p_{k} \leq P_t, \forall l}$. Let $p_i^*\brac{\bm{\lambda};\bm{p}^{(r)}}$, for $i=1,\ldots,K$, denote the transmit powers that minimize $\mathcal{L}\brac{\bm{p},\bm{\lambda};\bm{p}^{(r)}}$. As shown in Appendix~\ref{app:LangSoln}, it is given by
\begin{equation}
\label{Eq:PrimSoln}
p_i^*\brac{\bm{\lambda};\bm{p}^{(r)}}=p_i^{(r)}\sbrac{\frac{ A_i\brac{\bm{\lambda};\bm{p}^{(r)}}}{\mu_{j_i}+ B_i\brac{\bm{\lambda};\bm{p}^{(r)}}}}^+,
\end{equation}  
where $\mu_{j_i}$ is chosen to ensure that the power constraint is satisfied. Since $p_i^*\brac{\bm{\lambda};\bm{p}^{(r)}}$ is non-increasing in $\mu_{j_i}$, bisection method can be used to find in $\mu_{j_i}$. Here, $A_i\brac{\bm{\lambda};\bm{p}^{(r)}}$ and $B_i\brac{\bm{\lambda};\bm{p}^{(r)}}$ are given by
\begin{align}
\nonumber
&A_i\brac{\bm{\lambda};\bm{p}^{(r)}} = \sum_{k=1}^K \frac{\alpha_{k}\brac{w_{j_ik}+v\theta_{j_ik} s_{ik}}}{\sum_{n=1}^K  p_{n}^{(r)} \brac{w_{j_nk}+v\theta_{j_nk} s_{nk}} + \sigma^2} \\ \label{Eq:PowUpdNum}
& \hspace{0.5cm} + \frac{\lambda_{j_k}\brac{w_{j_ik}+v\theta_{j_ik} s_{ik} I_{\{i\neq k\}}}}{\sum_{n=1}^K  p_{n}^{(r)} \brac{w_{j_nk}+v\theta_{j_nk} s_{nk}} - v p_{k}^{(r)}\theta_{lk} + \sigma^2}, \\ \nonumber
& B_i\brac{\bm{\lambda};\bm{p}^{(r)}} = \sum_{k=1}^K \frac{\lambda_{j_k}\brac{w_{j_ik}+v\theta_{j_ik} s_{ik}}}{\sum_{n=1}^K  p_{n}^{(r)} \brac{w_{j_nk}+v\theta_{j_nk} s_{nk}} + \sigma^2} \\ \label{Eq:PowUpdDen}
& \hspace{0.5cm} + \frac{\alpha_{k} \brac{w_{j_ik}+v\theta_{j_ik} s_{ik} I_{\{i\neq k\}}}}{\sum_{n=1}^K  p_{n}^{(r)} \brac{w_{j_nk}+v\theta_{j_nk} s_{nk}} - v p_{k}^{(r)}\theta_{lk} + \sigma^2}.
\end{align}

The dual problem is now given by $\max_{\bm{\lambda}\geq \mathbf{0}}\;\mathcal{G}\brac{\bm{\lambda};\bm{p}^{(r)}}$, where $\mathcal{G}\brac{\bm{\lambda};\bm{p}^{(r)}}= \mathcal{L}\brac{\bm{p}^*\brac{\bm{\lambda};\bm{p}^{(r)}},\bm{\lambda};\bm{p}^{(r)}}$. It can be solved by gradient-ascent methods since $\mathcal{G}\brac{\bm{\lambda};\bm{p}^{(r)}}$ is a concave function of $\bm{\lambda}$~\cite{boyd_book}. Although $\mathcal{G}\brac{\bm{\lambda};\bm{p}^{(r)}}$ may not be differentiable, we can easily compute a subgradient $\mathbf{d}\brac{\bm{\lambda};\bm{p}^{(r)}}=\sbrac{d_1\brac{\bm{\lambda};\bm{p}^{(r)}},\ldots,d_L\brac{\bm{\lambda};\bm{p}^{(r)}}}$, where $d_l\brac{\bm{\lambda};\bm{p}^{(r)}}$, for $l=1,\ldots,L$, is given by~\cite{QYe_TCom_2016}
\begin{equation}
\label{Eq:Subgrad}
d_l\brac{\bm{\lambda};\bm{p}^{(r)}}=\sum_{k \in \mathcal{K}_l} H_{k}\brac{\bm{p}^*\brac{\bm{\lambda};\bm{p}^{(r)}};\bm{p}^{(r)}} - \tilde{\eta} C_{\text{plc}}.
\end{equation}
The dual variables are now updated as follows:
\begin{equation}
\label{Eq:DualUpd}
\bm{\lambda}_{t+1}=\sbrac{\bm{\lambda}_t+\delta_t \mathbf{d}\brac{\bm{\lambda}_t;\bm{p}^{(r)}}}^+,
\end{equation}
where $\delta_t$ is the step-size for iteration $t$. The dual variables $\bm{\lambda}_t$ converges to the optimal variables $\bm{\lambda}^*$ as $t \rightarrow \infty$. Since the duality gap for $\mathbf{P}2$ is zero and the dual solution is unique, the optimal solution to $\mathbf{P}2$ is given by $\bm{p}^*\brac{\bm{\lambda}^*;\bm{p}^{(r)}}$. 

In the $(r+1)\supth$ iteration of the SCA algorithm, the problem $\mathbf{P}2$ is solved with $\bm{p}^{(r)}$ replaced by $\bm{p}^{(r+1)}=\bm{p}^*\brac{\bm{\lambda}^*;\bm{p}^{(r)}}$. This is repeated until WSR converges. The overall SCA algorithm for WSR maximization is summarized in Algorithm~1.

\begin{algorithm}
\label{Alg:WSR}
\caption{WSR Maximization with Per-link Capacity Constraint}
Set $r=0$. Initialize with a feasible power vector $\bm{p}^{(r)}$

$\textbf{repeat}$ [SCA algorithm]

$\quad$ Set $t=0$ and $\bm{\lambda}_t=\mathbf{0}$

$\quad$ $\textbf{repeat}$ [dual subgradient algorithm]

$\quad$ $\quad$ Compute $\bm{p}^*\brac{\bm{\lambda}_t;\bm{p}^{(r)}}$ according to \eqref{Eq:PrimSoln}

$\quad$ $\quad$ Update $\bm{\lambda}_t$ according to \eqref{Eq:DualUpd} and $t=t+1$

$\quad$ $\textbf{until}$: relative change in $\mathcal{G}\brac{\bm{\lambda}_t;\bm{p}^{(r)}}$ is less than $\epsilon$

$\quad$ Update $\bm{p}^{(r+1)}= \bm{p}^*\brac{\bm{\lambda}^*;\bm{p}^{(r)}}$ and $r=r+1$

$\textbf{until}$: relative change in WSR is less than $\epsilon$
\end{algorithm}

We now provide an intuition for the power update in \eqref{Eq:PrimSoln}. For this, we first consider a simpler scenario with no fronthaul constraints. This corresponds to the case where the Lagrange multipliers are all zero, i.e., $\bm{\lambda}=\mathbf{0}$. Then, $A_i\brac{\mathbf{0};\bm{p}^{(r)}}$ and $B_i\brac{\mathbf{0};\bm{p}^{(r)}}$ from \eqref{Eq:PowUpdNum} and \eqref{Eq:PowUpdDen}, respectively, can be written as $A_i\brac{\mathbf{0};\bm{p}^{(r)}}=\sum_{k=1}^K \alpha_{k} \frac{d}{dp_i} U_k(\bm{p})\vert_{\bm{p^{(r)}}}$ and $B_i\brac{\mathbf{0};\bm{p}^{(r)}}=\sum_{k=1}^K \alpha_{k} \frac{d}{dp_i} V_k(\bm{p})\vert_{\bm{p^{(r)}}}$. For $k=0,\ldots,K$, $U_k(\bm{p})$ and $V_k(\bm{p})$ are  $U_k(\bm{p})=\log\brac{\sum_{n=1}^K  p_{n} \brac{w_{j_nk}+v\theta_{j_nk} s_{nk}} + \sigma^2}$ and $V_k(\bm{p}) = \log\left(\sum_{n=1}^K  p_{n} \brac{w_{j_nk}+v\theta_{j_nk} s_{nk}}  - v p_{k}\theta_{lk} + \sigma^2 \right)$.

Notice that $U_k(\bm{p})$ is the logarithm of the sum of signal, interference, and noise powers of user $k$ while $V_k(\bm{p})$ is the logarithm of the sum of its interference and noise powers. Further, $U_k(\bm{p})-V_k(\bm{p})$ gives the rate of user $k$ except for a scaling factor. Hence, we can identify $U_k(\bm{p})$ as the useful component in the rate expression for user $k$ as it involves the signal power and $V_k(\bm{p})$ as the harmful component. With this observation, we can see that power update in \eqref{Eq:PrimSoln} scales up or down the current power depending on whether the marginal change in the useful component aggregated over all the users exceed the marginal change in the harmful component again aggregated over all the users. This is quite intuitive. When the fronthaul constraints are present, additional penalty terms appear in the numerator and denominator of power update, which can again be understood as the marginal change in the useful and harmful components in the rate expression.

\emph{Convergence and Computational Complexity:} Since the bounds in Result~\ref{res:CvxApx} satisfy the properties in Lemma~\ref{Lem:SCA}, WSR monotonically increases and the power updates converge to a KKT point. The computational complexity of each SCA iteration is computed as follows\footnote{We ignore the computational complexity of bisection step in the analysis as it takes only a few iterations in general.}. The power and dual updates in \eqref{Eq:PrimSoln} and \eqref{Eq:DualUpd}, respectively, have $\mathcal{O}(K^3)$ complexity. Then, the complexity per SCA iteration is $\mathcal{O}(T_{\text{DSG}}K^3)$, where $T_{\text{DSG}}$ is the number of iterations needed for the subgradient algorithm to converge. It is dependent on the step size $\delta_t$ and the initialization point $\bm{\lambda}_0$. We have observed that the optimal dual variables for the previous iteration of the SCA algorithm is a good initialization point for the dual subgradient algorithm.

\subsection{Extension to Sum-capacity Constraint}

We now consider WSR maximization with sum-capacity constraint in \eqref{Eq:SC} instead of per-link capacity constraints in \eqref{Eq:PLC}. The approach is quite similar to before. In the $r\supth$ SCA iteration, we use the bounds in Result~\ref{res:CvxApx} to obtain a convex approximation of the original problem at $\bm{p}^{(r)}$, which is then solved by solving the corresponding dual program. Let $\lambda \geq 0$ be the Lagrange multiplier associated with the sum-capacity constraint. Then, the optimum power $p_i^*\brac{\lambda;\bm{p}^{(r)}}$ of user $i$, for $i=1,\ldots,K$, that maximizes the Lagrangian is given by
\begin{equation}
\label{Eq:PrimSolnSC}
p_i^*\brac{\lambda;\bm{p}^{(r)}}=p_i^{(r)}\sbrac{\frac{A_i\brac{\lambda\mathbf{1};\bm{p}^{(r)}}}{\mu_{j_i}+ B_i\brac{\lambda\mathbf{1};\bm{p}^{(r)}}}}^+,
\end{equation} 
where $\mu_{j_i}$ is chosen to satisfy the power constraint~\eqref{Eq:PowConstr} and can be efficiently computed via the bisection method. 

We use the dual subgradient algorithm to find the optimal dual variable $\lambda^*$ that maximizes the dual function. The dual update is computed similar to before and is given by
\begin{equation}
\label{Eq:DualUpdSC}
\!\lambda_{t+1} \! = \! \sbrac{\!\lambda_t \!+\! \delta_t \! \brac{\sum_{k =1}^K H_{k}\!\brac{\bm{p}^*\!\!\brac{\lambda_t;\bm{p}^{(r)}};\bm{p}^{(r)}\!} \!-\! \tilde{\eta} C_{\text{sc}}}\!}^+ \!\!. \!\!
\end{equation}

 Then, the solution to the convex problem in the $r\supth$ SCA iteration is $\bm{p}^*\brac{\lambda^*;\bm{p}^{(r)}}$. It is then used as the new point to evaluate the bounds in Result~\ref{res:CvxApx} in the next SCA iteration, i.e, $\bm{p}^{(r+1)}=\bm{p}^*\brac{\lambda^*;\bm{p}^{(r)}}$. The overall algorithm is summarized in Algorithm~2.

\begin{algorithm}
\label{Alg:WSRSC}
\caption{WSR Maximization for Sum-capacity Constraint}
Set $r=0$. Initialize with a feasible power vector $\bm{p}^{(r)}$

$\textbf{repeat}$ [SCA algorithm]

$\quad$ Set $t=0$ and $\lambda_t=0$

$\quad$ $\textbf{repeat}$ [dual subgradient algorithm]

$\quad$ $\quad$ Compute $\bm{p}^*\brac{\lambda_t;\bm{p}^{(r)}}$ according to \eqref{Eq:PrimSolnSC}

$\quad$ $\quad$ Update ${\lambda}_t$ according to \eqref{Eq:DualUpdSC} and $t=t+1$

$\quad$ $\textbf{until}$: relative change in dual objective is less than $\epsilon$

$\quad$ Update $\bm{p}^{(r+1)} = \bm{p}^*\brac{\lambda^*;\bm{p}^{(r)}}$ and $r=r+1$

$\textbf{until}$: relative change in WSR is less than $\epsilon$
\end{algorithm}

The convergence of the algorithm to a KKT point follows from Lemma~\ref{Lem:SCA}. The computational complexity of each iteration of the SCA algorithm is $\mathcal{O}(T_{\text{DSG}}K^3)$. Since the dual variable is a scalar, $T_\text{DSG}$ here is seen to be lower than that of the case with per-link capacity constraints.

\section{EE Maximization}
\label{sec:WEEMax}

We now focus on maximizing the EE of the C-RAN system. As before, we first formulate the optimization problem and develop an SCA algorithm for per-link capacity constraints and then for the sum-capacity constraint.

\subsection{Per-link Capacity Constraint}

Using the power consumption model in Section~\ref{sec:PowCons}, EE is defined as $\text{EE}=\sum_{i=1}^K R_i/P_{\text{C-RAN}}$. The EE maximization problem subject to transmit power constraints on RRUs and per-link capacity constraints on fronthaul is as follows:
\begin{equation}
\nonumber
\mathbf{P}3:\; \max_{\cbrac{\bm{p}\geq \mathbf{0}}} \; \text{EE}, \; \text{s.t.}\; \eqref{Eq:PowConstr},\;\eqref{Eq:PLC}.
\end{equation}
The above optimization problem is non-convex and global optimizers are hard to find. Therefore, we develop an SCA algorithm similar to WSR maximization.  

In order to apply the SCA algorithm, a lower bound on EE satisfying the properties in Lemma~\ref{Lem:SCA} is required. As shown in Appendix~\ref{App:EEBound}, such a lower bound, for any $\bm{p}$ and $\bm{p}^{(r)}$, is given by $ \text{EE} \geq \tau \sum_{i=1}^K G_{i}\brac{\bm{p};\bm{p}^{(r)}}/{P_{\text{C-RAN}}}$. Using this lower bound and the upper bound in \eqref{Eq:UB}, we get the following convex approximation for $\mathbf{P}3$ at $\bm{p}^{(r)}$ in the $r\supth$ SCA iteration:
\begin{equation}
\mathbf{P}4:\; \min_{\cbrac{\bm{p}\geq \mathbf{0}}}  - \frac{\sum_{i=1}^K G_{i}\brac{\bm{p};\bm{p}^{(r)}}}{P_S + \brac{\tau/\omega_{\text{RRU}}}\sum_{i=1}^K p_i}, \;\text{s.t.}\; \eqref{Eq:PowConstr},\;\eqref{Eq:PLCCvx}.
\nonumber
\end{equation}

Since $\mathbf{P}4$ involves the minimization of the ratio of a convex and an affine function over a convex set, its optimal solution can be found by the DB algorithm. In each iteration of it, the difference between the numerator term and the scaled denominator term of the objective function in $\mathbf{P}4$ is minimized over the same feasible set. The scaling factor for the denominator term is the value of the objective function in $\mathbf{P}4$ at the solution of the previous DB iteration. Thus, the equivalent problem solved in the $m\supth$ iteration of the DB algorithm is as follows:
\begin{equation}
\nonumber
\mathbf{P}5:\; \min_{\cbrac{\bm{p}\geq \mathbf{0}}} \; -\sum_{i=1}^K G_{i}\brac{\bm{p};\bm{p}^{(r)}} - \frac{q_m\tau}{\omega_{\text{RRU}}}\sum_{i=1}^K p_i, \;\text{s.t.}\; \eqref{Eq:PowConstr},\;\eqref{Eq:PLCCvx},
\end{equation}
where $q_m$ is given by
\begin{equation}
\label{Eq:DinkPar}
q_m=- \frac{\sum_{i=1}^K G_{i}\brac{\bm{p}^*\brac{q_{m-1},\bm{p}^{(r)}};\bm{p}^{(r)}}}{P_S + \brac{\tau/\omega_{\text{RRU}}}\sum_{i=1}^K p_i^*\brac{q_{m-1},\bm{p}^{(r)}}}.
\end{equation}
Here, $\bm{p}^*\brac{q_{m-1},\bm{p}^{(r)}}$ is the solution of $\mathbf{P}5$ in iteration $m-1$.

The optimizer of $\mathbf{P}5$ can be obtained by solving the dual problem in a manner similar to that in Section~\ref{sec:WSRMax}. Let $\bm{\lambda}=\lambda_1,\ldots,\lambda_L$ denote the Lagrange multipliers associated with the constraints in \eqref{Eq:PLCCvx}. Then, the optimal power $p_{i}^*\brac{\bm{\lambda};q_m,\bm{p}^{(r)}}$, for $i=1,\ldots,K$, that minimizes the Lagrangian function is
\begin{equation}
\label{Eq:PrimSolnEE}
p_{i}^*\brac{\bm{\lambda};q_m,\bm{p}^{(r)}}=\sbrac{\frac{p_i^{(r)} A_i\brac{\bm{\lambda};\bm{p}^{(r)}}}{\mu_{j_i}-\frac{q_m\tau}{\omega_{\text{RRU}}}+ B_i\brac{\bm{\lambda};\bm{p}^{(r)}}}}^+,
\end{equation}  
where $\mu_{j_i}$ is chosen to ensure that the power constraint \eqref{Eq:PowConstr} is satisfied and can computed using the bisection method. The dual variables are updated as per \eqref{Eq:DualUpd}. Let $\bm{\lambda}^*$ denote the optimal dual variable. Then, the solution to $\mathbf{P}5$ is given by $\bm{p}^*\brac{\bm{\lambda}^*;q_m,\bm{p}^{(r)}}$. It is then used to compute $q_{m+1}$ using \eqref{Eq:DinkPar}. This is then repeated until the sequence of $q_m$ converges and let $q^*$ be its limit. Then, the solution to $\mathbf{P}4$ is given by $\bm{p}^*\brac{\bm{\lambda}^*;q^*,\bm{p}^{(r)}}$. It is then used as $\bm{p}^{(r+1)}$ for the next SCA iteration. The overall algorithm is summarized in Algorithm~3.

\begin{algorithm}
\label{Alg:EEPLC}
\caption{EE Maximization with Per-link Capacity Constraint}
Set $r=0$. Initialize with a feasible power vector $\bm{p}^{(r)}$

$\textbf{repeat}$ [SCA algorithm]

$\quad$ Set $m=0$. Initialize $q_m=0$

$\quad$ $\textbf{repeat}$ [DB algorithm]

$\quad$ Set $t=0$ and $\bm{\lambda}_t=\mathbf{0}$

$\quad$ $\quad$ $\textbf{repeat}$ [Dual subgradient algorithm]

$\quad$ $\quad$ $\quad$ Compute $\bm{p}^*\brac{\bm{\lambda}_t;q_m,\bm{p}^{(r)}}$ using \eqref{Eq:PrimSolnEE}

$\quad$ $\quad$ $\quad$ Update $\bm{\lambda}_t$ according to \eqref{Eq:DualUpd} and $t=t+1$

$\quad$ $\quad$ $\textbf{until}$: relative change in dual objective is below $\epsilon$

$\quad$ $\quad$ Update $q_m$ using \eqref{Eq:DinkPar} and $m=m+1$

$\quad$ $\textbf{until}$: relative change in the objective function in $\mathbf{P}4$ is less than $\epsilon$

$\quad$ Update $\bm{p}^{(r)}=\bm{p}^*\brac{\bm{\lambda}^*;q^*,\bm{p}^{(r)}}$ and  $r=r+1$

$\textbf{until}$: relative change in EE is less than $\epsilon$
\end{algorithm}

\emph{Convergence and Complexity:} The convergence of the above algorithm to a KKT point is ensured by Lemma~\ref{Lem:SCA}. We note that the DB algorithm has been employed for EE maximization in~\cite{Schober_TWC_2012}. However, they apply the DB algorithm to the original non-convex problem unlike our approach. Hence, the convergence of the DB algorithm is not guaranteed in~\cite{Schober_TWC_2012} as they do not ensure that the global optimizer is found in each iteration of the DB algorithm. This issue does not arise in our approach since convex problems are solved in each iteration of the DB algorithm.

The computational complexity of each iteration of the DB algorithm is $\mathcal{O}(T_{\text{DSG}}K^3)$ as it is similar to $\mathbf{P}2$. Let $T_{\text{DB}}$ be the number of iterations of the DB algorithm. Then, the overall complexity of each iteration of the SCA algorithm is $\mathcal{O}(T_{\text{DB}}T_{\text{DSG}}K^3)$. Since the DB algorithm has super-linear convergence~\cite{Schober_TWC_2012}, $T_{\text{DB}}$ is quite small.

\subsection{Extension to Sum-capacity Constraint}

We now consider EE maximization with sum-capacity constraint. This problem is similar to $\mathbf{P}3$ except that the $L$ constraints in \eqref{Eq:PLC} are replaced by a single constraint in \eqref{Eq:SC}. The optimization problem is solved in a manner similar to above. The only changes are the solution to the Lagrangian function and dual problem in each DB iteration. These changes arise as there is just one fronthaul constraint unlike before. Let $\lambda\geq 0$ denote the Lagrangian multiplier associated with the fronthaul constraint in \eqref{Eq:SC}. Then, the optimal power $p_{i}^*\brac{\lambda;q_m,\bm{p}^{(r)}}$, for $i=1,\ldots,K$, that minimizes the Lagrangian function is
\begin{equation}
\label{Eq:PrimSolnEESC}
p_{i}^*\brac{\lambda;q_m,\bm{p}^{(r)}}=\sbrac{\frac{p_i^{(r)} A_i\brac{\lambda\mathbf{1};\bm{p}^{(r)}}}{\mu_{j_i}-\frac{q_m\tau}{\omega_{\text{RRU}}}+ B_i\brac{\lambda \mathbf{1};\bm{p}^{(r)}}}}^+,
\end{equation}  
where $\mu_{j_i}$ is chosen to ensure that the power constraint in \eqref{Eq:PowConstr} is satisfied and can be efficiently computed using the bisection method. The optimal dual variable $\lambda^*$, which maximizes the dual function, is obtained via the dual subgradient algorithm. The overall algorithm is summarized in Algorithm~4.

\begin{algorithm}
\label{Alg:EESC}
\caption{EE Maximization with Sum-capacity Constraint}
Set $r=0$. Initialize with a feasible power vector $\bm{p}^{(r)}$

$\textbf{repeat}$ [SCA algorithm]

$\quad$ Set $m=0$. Initialize $q_m=0$

$\quad$ $\textbf{repeat}$ [DB algorithm]

$\quad$ Set $t=0$ and $\lambda_t=\mathbf{0}$

$\quad$ $\quad$ $\textbf{repeat}$ [Dual subgradient algorithm]

$\quad$ $\quad$ $\quad$ Compute $\bm{p}^*\brac{{\lambda}_t;q_m,\bm{p}^{(r)}}$ using \eqref{Eq:PrimSolnEESC}

$\quad$ $\quad$ $\quad$ Update $\lambda_t$ according to \eqref{Eq:DualUpdSC} and $t=t+1$

$\quad$ $\quad$ $\textbf{until}$: relative change in dual objective is below $\epsilon$

$\quad$ $\quad$ Update $q_m$ using \eqref{Eq:DinkPar} and $m=m+1$

$\quad$ $\textbf{until}$: relative change in the objective function in $\mathbf{P}4$ is less than $\epsilon$

$\quad$ Update $\bm{p}^{(r)}=\bm{p}^*\brac{\lambda^*;q^*,\bm{p}^{(r)}}$ and $r=r+1$

$\textbf{until}$: relative change in EE is less than $\epsilon$
\end{algorithm}

The convergence of Algorithm~4 to a KKT point is ensured by Lemma~\ref{Lem:SCA}. The computational complexity of each SCA iteration can be computed as before and is $\mathcal{O}(T_{\text{DB}}T_{\text{DSG}}K^3)$. 

\section{Numerical Results}
\label{sec:NumRes}

We now carry out Monte Carlo simulations to evaluate the performance of the proposed SCA algorithms. We consider a 7-cell hexagonal cellular layout with wrap-around. The cell radius is set to $500$ m. We drop $K=70$ users randomly in the network area. The large-scale fading coefficients are computed as follows. The pathloss in dB at distance $d$ in km is $128.1+3.76\log_{10}(d)$. The standard deviation of lognormal shadowing is $8$~dB. The other simulation parameters are $N=200$, $T_{\text{coh}}=200$, $T_p=10$, $P_{\text{tr}} =23$~dBm, $P_{\text{dl}} =46$~dBm, $\varsigma=0.2$~W, $\varrho=1.8$~W, $\omega_\text{RRU}=\omega_\text{UE}=0.3$, $\kappa=\eta=1$, and $\epsilon=0.01$. The signal bandwidth is $10$~MHz. The pilot sequences are reused in each cell and they are allocated randomly to the users.

We consider distance-based and signal power-based~\cite{Bjornson_TWC_2016} association rules. In the former, a user associates to the RRU that is closest to it. In the latter, a user associates to the RRU from which it receives the highest signal power when RRUs are transmitting at full power. Thus, the user association here depends on both pathloss and shadowing. We note that it is not straightforward to implement the commonly considered \emph{max-SINR} rule, in which a user associates to the RRU that provides it the highest SINR. This is because SINR depends on pilot allocation, which, in turn, depends on user association.

We benchmark the proposed WSR maximization algorithms against the WMMSE algorithm in~\cite{WeiYu_Access_2014}. While the WMMSE algorithm can not be applied directly to our setting, it can be adapted by appropriately redefining mean square error. For EE benchmarking, we use an extension of the WMMSE algorithm. We employ the DB algorithm as used in~\cite{Schober_TWC_2012} to extend the WMMSE algorithm to handle EE maximization. We note that these WMMSE algorithms lacks any convergence guarantees. Therefore, we limit the maximum number of iterations to $100$. We also benchmark the proposed algorithms against a baseline scheme in which there is no coordination between the RRUs. In it, the RRUs transmit at equal power and the power value is chosen to ensure that the fronthaul constraints are satisfied. 

We first present the results for WSR maximization and then for EE maximization. For WSR maximization, we set the user weights to be equal, i.e., the users have equal priority. 

\subsection{Network Throughput}

\begin{figure*}
\centering
\subfloat[MRT, Per-link capacity constraint\label{Fig:SR_MRC_PLC}]{\includegraphics[width=0.3\linewidth]{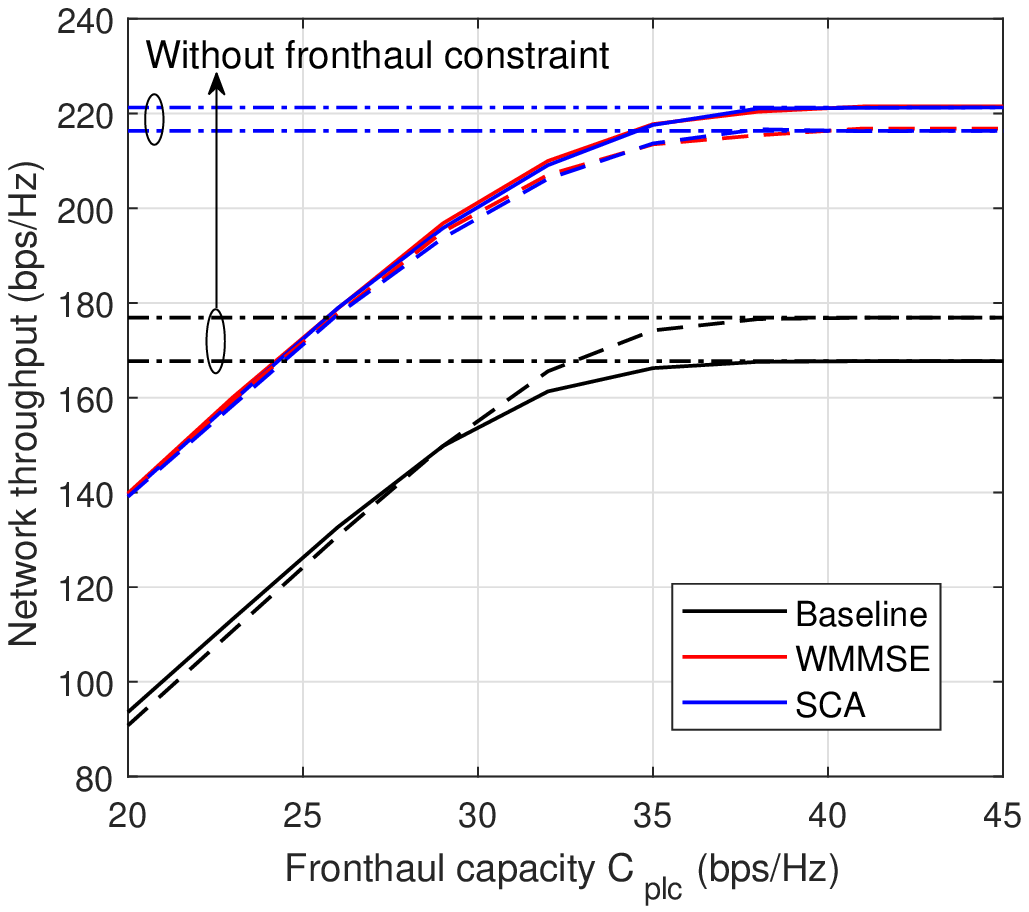}}
\subfloat[ZF, Per-link capacity constraint\label{Fig:SR_ZF_PLC}] {\includegraphics[width=0.3\linewidth]{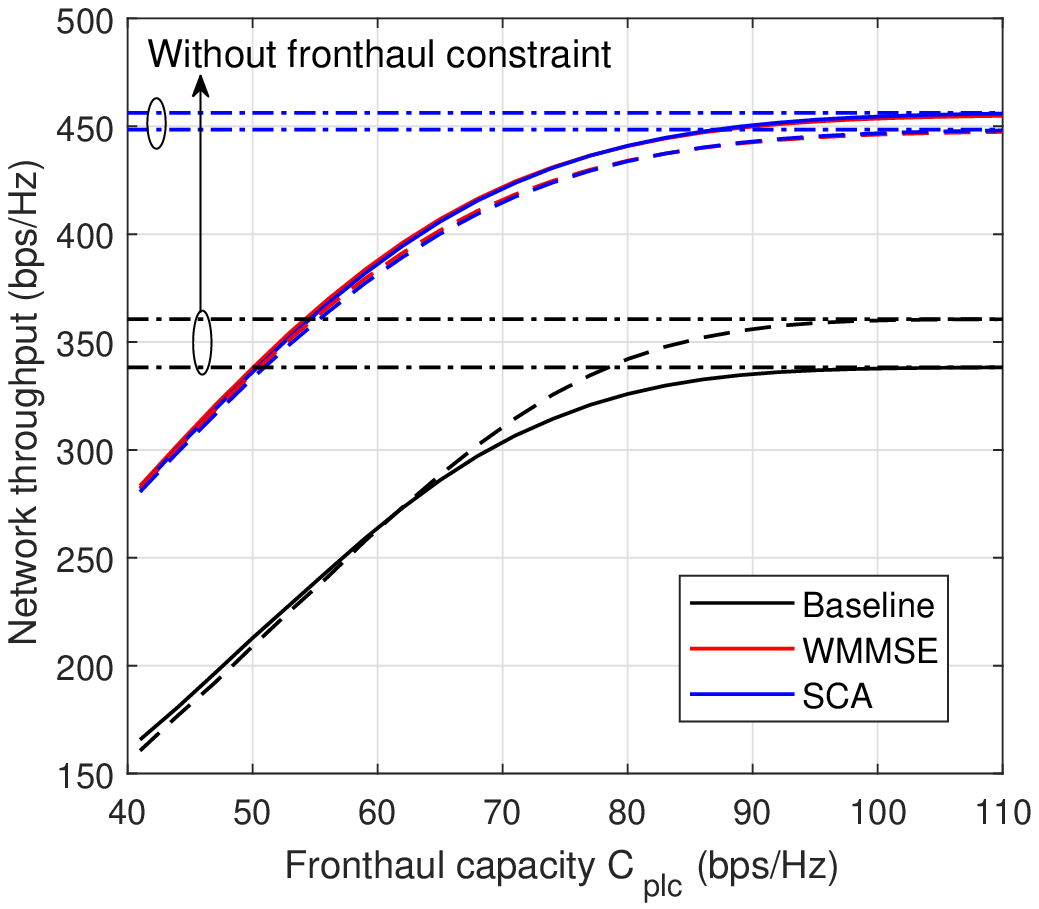}}
\subfloat[ZF, Sum-capacity constraint\label{Fig:SR_ZF_SC}]{\includegraphics[width=0.3\linewidth]{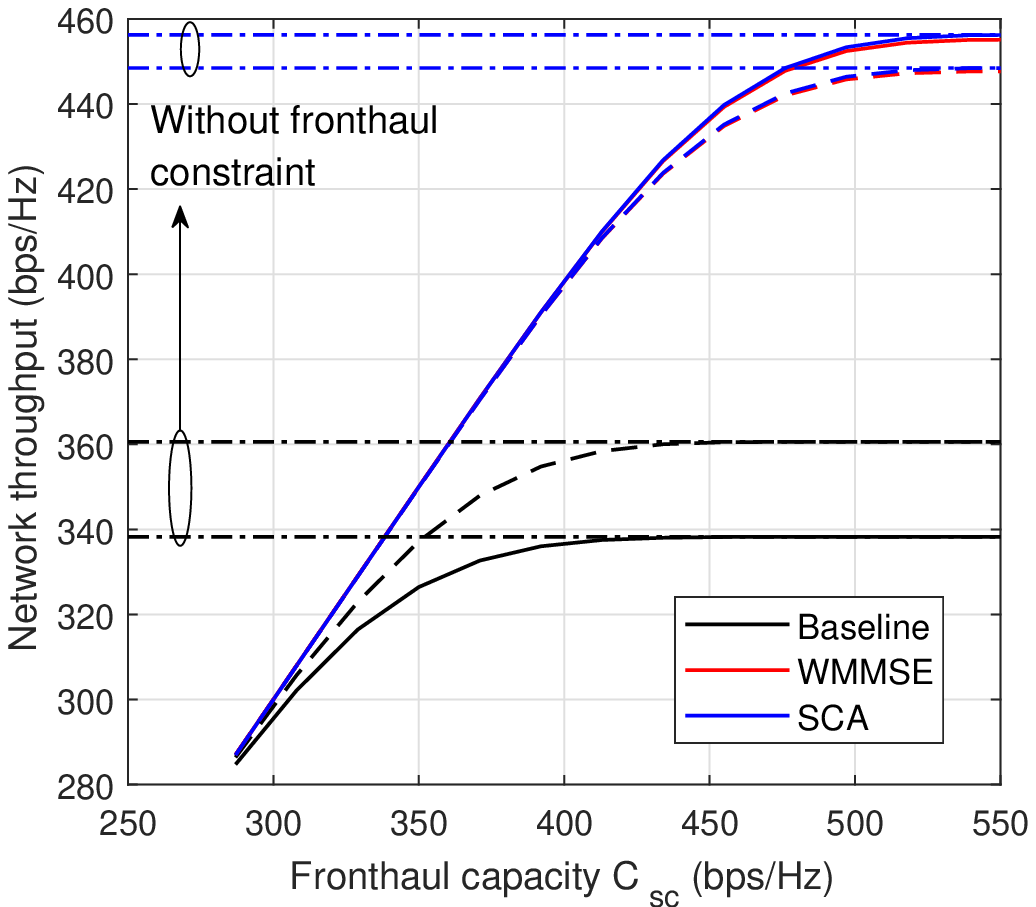}}
\caption{Zoomed-in view of network throughput against fronthaul capacity for different schemes. The solid curves denote signal power-based association and dashed curves denote distance-based association.}
\end{figure*}

Figures~\ref{Fig:SR_MRC_PLC} and \ref{Fig:SR_ZF_PLC} show the results for network throughput, which is the sum rate of the users averaged over several channel realizations, with per-link capacity constraints for MRT and ZF precoding, respectively. We see that the SCA algorithm achieves a significantly higher network throughput than the baseline scheme for both the association rules. For example, the SCA algorithm improves the throughput by $54\%$, $29\%$, and $25\%$ at $C_{\text{plc}}=20$, $30$, and $40$~bps/Hz, respectively, with signal power-based association and MRT precoding. The corresponding gains for ZF precoding at $C_{\text{plc}}=50$, $70$, and $90$~bps/Hz are $59\%$, $38\%$, and $34\%$, respectively. The WMMSE algorithm achieves a similar performance as the SCA algorithm for both the precoding schemes. 

We see that the throughput trends are similar for the proposed algorithm and the benchmark schemes. The network throughput increases as fronthaul capacity increases before saturating. This is because the fronthaul constraints become more relaxed as fronthaul capacity increases. Eventually, the throughput is determined only by the power constraints. This is evident from the figures as the curves approach the throughput curves for the case without fronthaul constraints.

We also see that ZF precoding achieves a significantly higher throughput than MRT precoding for all the schemes. This is because the former manages the interference better than the latter. We also see that the curves for the two association rules are close to each other for both SCA and WMMSE algorithms. For example, the maximum difference is less than $2.5\%$. However, this is not the case for the baseline scheme. Thus, with power control between the RRUs, the initial user association does not seem to impact the performance much.

Figure \ref{Fig:SR_ZF_SC} shows the network throughput results for ZF precoding with sum-capacity constraint. As before, we see that the SCA and WMMSE algorithms significantly improve the network throughput over the baseline scheme. However, at very low fronthaul capacities, all three schemes have the same throughput, which equals the sum-capacity of fronthaul since this is an upper bound on the throughput. The difference in throughputs for the two association rules is again small for SCA and WMMSE algorithms and is less than $2\%$. The results are similar for MRT precoding and are skipped to save space.

We now study the impact of the nature of the fronthaul constraint on network throughput. For this, we compare the network throughput with per-link constraint at $C_{\text{plc}}$ and network throughput with sum-capacity constraint at $C_{\text{sc}}=LC_{\text{plc}}$. The reason for it is that the fronthaul link between BBU and switch carries data to $L$ RRUs. We see that the throughput is higher for the case with sum-capacity constraint. This is intuitive as sum-capacity constraint is a more relaxed constraint than per-link capacity constraints. Thus, the impact on the throughput is lower if the fronthaul link between BBU and switch is the bottleneck rather than the links between switch and RRUs.

\subsection{Energy Efficiency}

\begin{figure*}
\vspace{-0.4cm}
\centering
\subfloat[MRT, Per-link capacity constraint\label{Fig:EE_MRC_PLC}]{\includegraphics[width=0.3\linewidth]{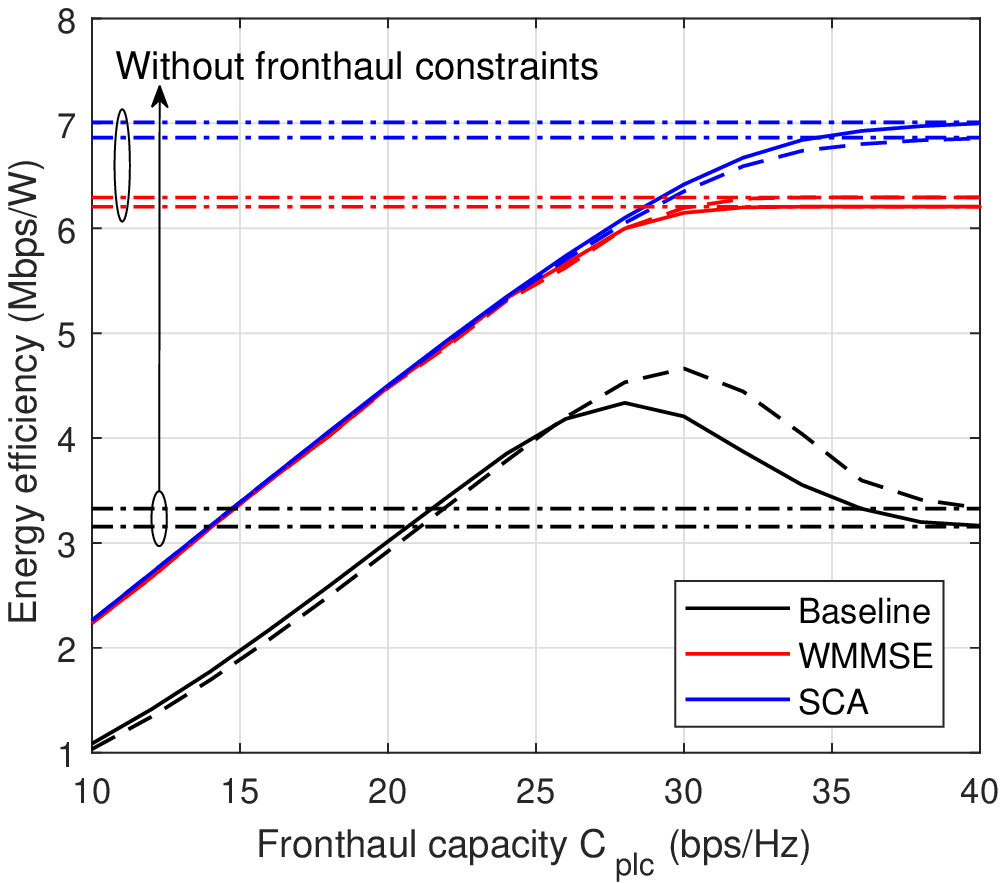}}
\subfloat[ZF, Per-link capacity constraint\label{Fig:EE_ZF_PLC}] {\includegraphics[width=0.3\linewidth]{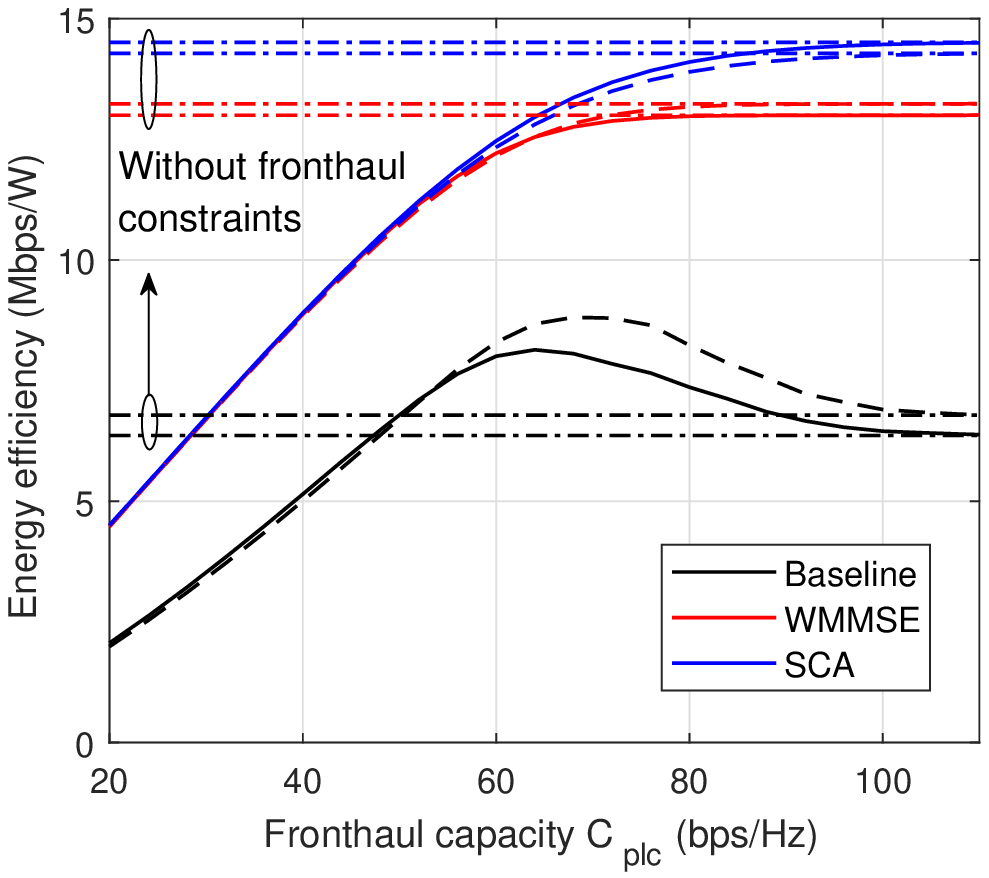}}
\subfloat[MRT, Sum-capacity constraint\label{Fig:EE_ZF_SC}]{\includegraphics[width=0.3\linewidth]{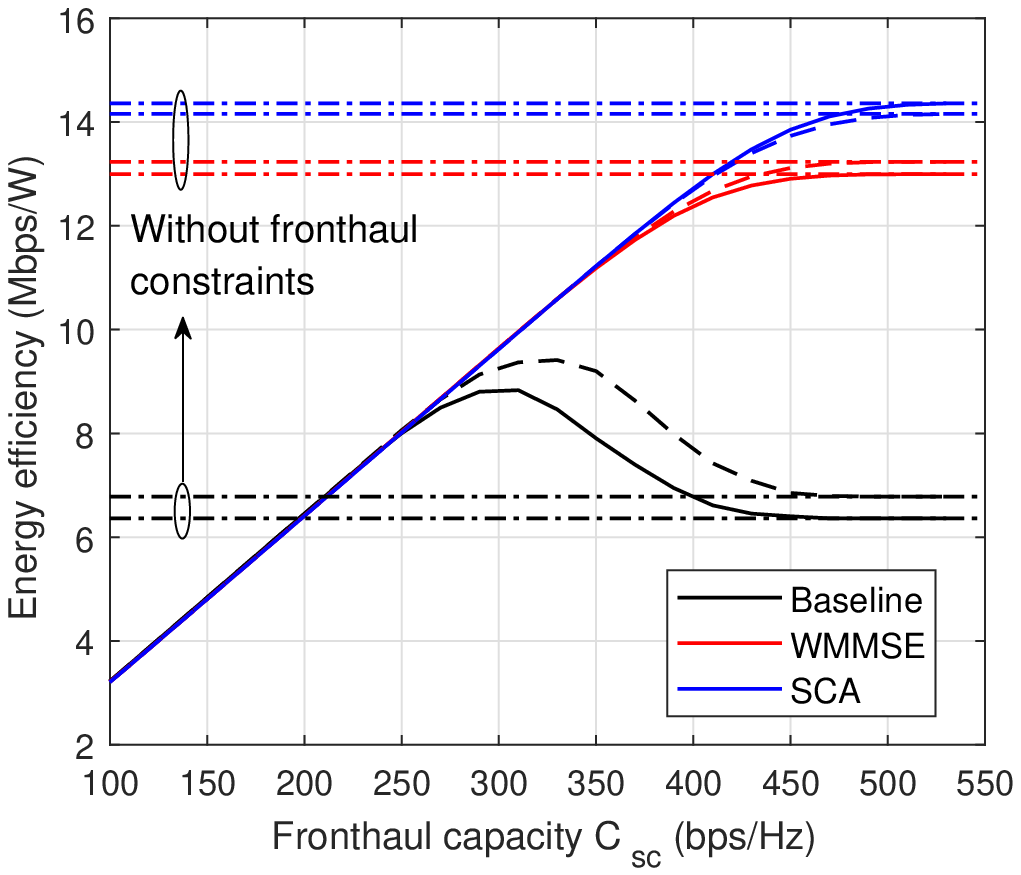}}
\caption{Zoomed-in view of EE against fronthaul capacity for different schemes. The solid curves denote signal power-based association and dashed curves denote distance-based association.} 
\end{figure*}

Figures~\ref{Fig:EE_MRC_PLC} and \ref{Fig:EE_ZF_PLC} show the EE results with per-link capacity constraints for MRT and ZF precoding, respectively. We see that the SCA algorithm achieves a significantly higher EE than the baseline scheme. For example, the SCA algorithm improves EE by $49\%$, $53\%$, and $122\%$ at $C_{\text{plc}}=20$, $30$, and $40$~bps/Hz, respectively, with signal power-based association and MRT precoding. The gains for ZF precoding at $C_{\text{plc}}=40$, $60$, and $80$~bps/Hz are $73\%$, $56\%$, and $91\%$, respectively. While SCA and WMMSE algorithms have similar performance at low values of $C_{\text{plc}}$, the SCA algorithm outperforms the WMMSE algorithm at high values of $C_{\text{plc}}$. For example, the SCA algorithm improves EE by $12\%$ at $C_{\text{plc}}=100$~bps/Hz over the WMMSE algorithm for ZF precoding. 

For SCA and WMMSE algorithms, we see that EE increases as $C_{\text{plc}}$ increases before saturating. However, for the baseline scheme, EE first increases and then decreases. This is because the RRUs transmit at higher powers as fronthaul capacity increases. Beyond a certain point, however, this increase in transmit power does not translate to an increase in sum rate. This causes the EE to decrease. We also see that the difference in EE with the two association rules is more pronounced for the baseline scheme than SCA and WMMSE algorithms. 

Figure~\ref{Fig:EE_ZF_SC} shows the EE results for ZF precoding with sum-capacity constraint. We see that the trends are quite similar to those with the per-link capacity constraint. Similar to the case of WSR maximization, we see that the baseline scheme achieves the same performance as that of SCA and WMMSE algorithms at low fronthaul capacities. However, this is not the case at high fronthaul capacities.

\section{Conclusions}
\label{sec:Concls}

We developed an optimization framework based on SCA to optimize the transmit powers of the users to maximize WSR and EE. The framework explicitly incorporated the capacity constraint on fronthaul. We considered per-link and sum capacity constraints, which respectively modeled the capacity constraint on links to RRUs and the link between BBU and switch. We derived novel bounds on the non-convex user rate function to apply the SCA algorithm. These bounds yielded convex problems in each iteration of the SCA algorithm, which were solved by solving their dual problems. We saw that the proposed algorithms significantly improved network throughput and EE over a baseline scheme. They also exhibited performance similar to or better than the WMMSE algorithm.

\section{Appendix}

\subsection{Brief Introduction to SCA}
\label{app:SCA}
 
The idea behind SCA is to tackle a difficult non-convex problem by solving a sequence of easier convex problems. The following is a key result for SCA; refer~\cite{Alessio_EE_Book} for proof.

\begin{lemma}
\label{Lem:SCA}
Let $\mathbf{P}$ be a minimization problem with continuous objective $f_0(\mathbf{x})$ and constraints $f_i(\mathbf{x}) \leq 0$, for $i=1,\ldots,m$, defining a compact set. Let $\cbrac{\mathbf{P}_l}_l$ be a sequence of minimization problems with objective $f_{0,l}(\mathbf{x})$, constraints $f_{i,l}(\mathbf{x}) \leq 0$, for $i=1,\ldots,m$, and optimal solution $\mathbf{x}_l^*$. Assume that $f_{i,l}(\mathbf{x})$, for all $l$ and $i=0,\ldots,m$, satisfies the following properties: 1) $f_{i}(\mathbf{x}) \leq f_{i,l}(\mathbf{x})$, for all $\mathbf{x}$, 2) $f_{i}(\mathbf{x}_{l-1}^*)=f_{i,l}(\mathbf{x}_{l-1}^*)$, 3) $\nabla f_{i}(\mathbf{x}_{l-1}^*)= \nabla f_{i,l}(\mathbf{x}_{l-1}^*)$. Then, the sequence $\cbrac{\mathbf{x}_l^*}_l$ converges monotonically to a point satisfying the KKT conditions of the original problem $\mathbf{P}$.
\end{lemma}

\subsection{Derivation of $G_{k}\brac{\bm{p};\bm{p}^{(r)}}$ and $H_{k}\brac{\bm{p};\bm{p}^{(r)}}$}
\label{app:Bounds}

We first show the derivation of $H_{k}\brac{\bm{p};\bm{p}^{(r)}}$ and then for $H_{k}\brac{\bm{p};\bm{p}^{(r)}}$. From Section~\ref{sec:WSRMax}, we have $\log\brac{1+\gamma_{k}}=U_k(\bm{p})-V_k(\bm{p})$. We now provide an upper bound for $U_k(\bm{p})$ and a lower bound for $V_k(\bm{p})$. 

Since $U_k(\bm{p})$ is concave in $\bm{p}$, for any $\bm{p}^{(r)}$, we have
\begin{equation}
\label{Eq:UBTm1}
U_k(\bm{p}) \leq U_k\brac{\bm{p}^{(r)}}+ \sum_{i=1}^K (p_i-p_i^{(r)}) \frac{d}{dp_i} U_k(\bm{p}) \vert_{\bm{p}^{(r)}}.
\end{equation}
For $V_k(\bm{p})$, we use the variable transformation $\tilde{p}_i=\log(p_i)$, for $i=1,\ldots,K$. Since $V_k(\tilde{\bm{p}})$ is the composition of affine and log-sum-exp functions, it is convex in $\tilde{\bm{p}}$~\cite{boyd_book}. Thus, for any $\tilde{\bm{p}}^{(r)}$, we have
\begin{equation}
\label{Eq:UBTm2}
V_k(\tilde{\bm{p}}) \geq V_k\brac{\tilde{\bm{p}}^{(r)}}+ \sum_{i=1}^K (\tilde{p}_i-\tilde{p}_i^{(r)}) \frac{d}{d\tilde{p}_i} V_k(\tilde{\bm{p}}) \vert_{\tilde{\bm{p}}^{(r)}}.
\end{equation}
The derivatives in \eqref{Eq:UBTm1} and \eqref{Eq:UBTm2} can be easily evaluated. We then transform $\tilde{\bm{p}}$ in \eqref{Eq:UBTm2} back to $\bm{p}$. Thereafter, the difference between bounds in 
\eqref{Eq:UBTm1} and \eqref{Eq:UBTm2} yields $H_{k}\brac{\bm{p};\bm{p}^{(r)}}$ in \eqref{Eq:UB}. Since this bound follows from concavity of $U_k(\bm{p})$ and convexity of $V_k(\tilde{\bm{p}})$, it satisfies the properties in Lemma~\ref{Lem:SCA}.

In order to derive $G_{k}\brac{\bm{p};\bm{p}^{(r)}}$, a lower bound on $U_k(\bm{p})$ and an upper bound on $V_k(\bm{p})$ is obtained in a manner similar to above. We skip the details to conserve space. 

\subsection{Solution of Lagrangian Minimization}
\label{app:LangSoln}

In order to solve $\min_{\mathcal{P}} \mathcal{L}\brac{\bm{p},\bm{\lambda};\bm{p}^{(r)}}$, we use the KKT conditions, which are necessary and sufficient in this case. They are $
\frac{d}{dp_i} \sbrac{ \mathcal{L}\brac{\bm{p},\bm{\lambda};\bm{p}^{(r)}} + \sum_{k=1}^K \mu_{j_k} p_{k} - \zeta_k p_k } =0$, $\mu_l\brac{\sum_{k \in \mathcal{K}_l} p_{k}-P_t}=0$, $\zeta_np_n=0$, $\sum_{k \in \mathcal{K}_l} p_{k} \leq P_t$, $p_k \geq 0$, for $l=1,\ldots,L$ and $i,k=1,\ldots,K$. Here, $\mu_l\geq 0$ and $\zeta_n\geq 0$ are the KKT multipliers. 

We get $p_i^*\brac{\bm{\lambda};\bm{p}^{(r)}}$ by solving the derivative condition. This coupled with the conditions $\zeta_np_n=0,\;p_n \geq 0,\;\zeta_n \geq 0$ yields \eqref{Eq:PrimSoln}. The KKT multiplier $\mu_l$ is chosen such that $\sum_{k \in \mathcal{K}_l} p_{k} = P_t$ and is set to zero if it is negative. This choice of $\mu_l$ satisfies the KKT conditions relating to $\mu_l$.

%

\subsection{Lower Bound on EE}
\label{App:EEBound}

A lower bound on EE is obtained by replacing $\log\brac{1+\gamma_{k}}$ with $G_{k}\brac{\bm{p};\bm{p}^{(r)}}$. We now show that the resulting lower bound satisfies the properties in Lemma~\ref{Lem:SCA}. It is easy to see that the first two properties are satisfied. For the third property, we need to show that $ \frac{d}{dp_i} \text{EE} \rvert_{\bm{p}^{(r)}} = \frac{d}{dp_i} \frac{\sum_{k=1}^K G_{k}\brac{\bm{p};\bm{p}^{(r)}}}{P_{\text{C-RAN}}} \rvert_{\bm{p}^{(r)}}$. Let $c=[P_S+ (\tau/\omega_{\text{RRU}})\sum_{i=1}^K p_i^{(r)}]^{-1}$. Then, we need to show 
\begin{multline}
\nonumber
{\sum_{k=1}^K \frac{d}{dp_i} \log\brac{1+\gamma_{k}} \rvert_{\bm{p}^{(r)}}} -\brac{\tau c/ \omega_{\text{RRU}}} \log\brac{1+\gamma_{k}^{(r)}} \\
 = \sum_{k=1}^K \frac{d}{dp_i} G_{k}\brac{\bm{p};\bm{p}^{(r)}} \rvert_{\bm{p}^{(r)}} - \brac{\tau c/ \omega_{\text{RRU}}} G_{k}\brac{\bm{p}^{(r)};\bm{p}^{(r)}}.
\end{multline}
%
It follows as $G_{k}\brac{\bm{p};\bm{p}^{(r)}}$ satisfies the properties in Lemma~\ref{Lem:SCA}.


\end{document}